\documentclass[useAMS,usenatbib]{mn2e}
\pdfoutput=1
\usepackage{natbib}
\usepackage{graphicx}
\usepackage{amssymb}
\usepackage{lineno}
\usepackage{amsmath}
\begin{document}
%%%%%%%%%%%%%%%%%%%%%%%%%%%%%%%%%%%%%%%
\newcommand{\MSun}{{M_\odot}}
\newcommand{\LSun}{{L_\odot}}
\newcommand{\Rstar}{{R_\star}}
\newcommand{\calE}{{\cal{E}}}
\newcommand{\calM}{{\cal{M}}}
\newcommand{\calV}{{\cal{V}}}
\newcommand{\calO}{{\cal{O}}}
\newcommand{\calH}{{\cal{H}}}
\newcommand{\calD}{{\cal{D}}}
\newcommand{\calB}{{\cal{B}}}
\newcommand{\calK}{{\cal{K}}}
\newcommand{\labeln}[1]{\label{#1}}
\newcommand{\Lsolar}{L$_{\odot}$}
\newcommand{\farcmin}{\hbox{$.\mkern-4mu^\prime$}}
\newcommand{\farcsec}{\hbox{$.\!\!^{\prime\prime}$}}
\newcommand{\kms}{\rm km\,s^{-1}}
\newcommand{\cc}{\rm cm^{-3}}
\newcommand{\Alfven}{$\rm Alfv\acute{e}n$}
\newcommand{\Vap}{V^\mathrm{P}_\mathrm{A}}
\newcommand{\Vat}{V^\mathrm{T}_\mathrm{A}}
\newcommand{\D}{\partial}
\newcommand{\DD}{\frac}
\newcommand{\TAW}{\tiny{\rm TAW}}
\newcommand{\mm }{\mathrm}
\newcommand{\Bp }{B_\mathrm{p}}
\newcommand{\Bpr }{B_\mathrm{r}}
\newcommand{\Bpz }{B_\mathrm{\theta}}
\newcommand{\Bt }{B_\mathrm{T}}
\newcommand{\Vp }{V_\mathrm{p}}
\newcommand{\Vpr }{V_\mathrm{r}}
\newcommand{\Vpz }{V_\mathrm{\theta}}
\newcommand{\Vt }{V_\mathrm{\varphi}}
\newcommand{\Ti }{T_\mathrm{i}}
\newcommand{\Te }{T_\mathrm{e}}
\newcommand{\rtr }{r_\mathrm{tr}}
\newcommand{\rbl }{r_\mathrm{BL}}
\newcommand{\rtrun }{r_\mathrm{trun}}
\newcommand{\thet }{\theta}
\newcommand{\thetd }{\theta_\mathrm{d}}
\newcommand{\thd }{\theta_d}
\newcommand{\thw }{\theta_W}
\newcommand{\beq}{\begin{equation}}
\newcommand{\eeq}{\end{equation}}
\newcommand{\ben}{\begin{enumerate}}
\newcommand{\een}{\end{enumerate}}
\newcommand{\bit}{\begin{itemize}}
\newcommand{\eit}{\end{itemize}}
\newcommand{\barr}{\begin{array}}
\newcommand{\earr}{\end{array}}
\newcommand{\bc}{\begin{center}}
\newcommand{\ec}{\end{center}}
\newcommand{\DroII}{\overline{\overline{\rm D}}}
\newcommand{\DroI}{{\overline{\rm D}}}
\newcommand{\eps}{\epsilon}
\newcommand{\veps}{\varepsilon}
\newcommand{\vepsdi}{{\cal E}^\mathrm{d}_\mathrm{i}}
\newcommand{\vepsde}{{\cal E}^\mathrm{d}_\mathrm{e}}
\newcommand{\lraS}{\longmapsto}
\newcommand{\lra}{\longrightarrow}
\newcommand{\LRA}{\Longrightarrow}
\newcommand{\Equival}{\Longleftrightarrow}
\newcommand{\DRA}{\Downarrow}
\newcommand{\LLRA}{\Longleftrightarrow}
\newcommand{\diver}{\mbox{\,div}}
\newcommand{\grad}{\mbox{\,grad}}
\newcommand{\cd}{\!\cdot\!}
\newcommand{\Msun}{{\,{\cal M}_{\odot}}}
\newcommand{\Mstar}{{\,{\cal M}_{\star}}}
\newcommand{\Mdot}{{\,\dot{\cal M}}}
\newcommand{\ds}{ds}
\newcommand{\dt}{dt}
\newcommand{\dx}{dx}
\newcommand{\dr}{dr}
\newcommand{\dth}{d\theta}
\newcommand{\dphi}{d\phi}

\newcommand{\pt}{\frac{\partial}{\partial t}}
\newcommand{\pk}{\frac{\partial}{\partial x^k}}
\newcommand{\pj}{\frac{\partial}{\partial x^j}}
\newcommand{\pmu}{\frac{\partial}{\partial x^\mu}}
\newcommand{\pr}{\frac{\partial}{\partial r}}
\newcommand{\pth}{\frac{\partial}{\partial \theta}}
\newcommand{\pR}{\frac{\partial}{\partial R}}
\newcommand{\pZ}{\frac{\partial}{\partial Z}}
\newcommand{\pphi}{\frac{\partial}{\partial \phi}}

\newcommand{\vadve}{v^k-\frac{1}{\alpha}\beta^k}
\newcommand{\vadv}{v_{Adv}^k}
\newcommand{\dv}{\sqrt{-g}}
\newcommand{\fdv}{\frac{1}{\dv}}
\newcommand{\dvr}{{\tilde{\rho}}^2\sin\theta}
\newcommand{\dvt}{{\tilde{\rho}}\sin\theta}
\newcommand{\dvrss}{r^2\sin\theta}
\newcommand{\dvtss}{r\sin\theta}
\newcommand{\dd}{\sqrt{\gamma}}
\newcommand{\ddw}{\tilde{\rho}^2\sin\theta}
\newcommand{\mbh}{M_{BH}}
\newcommand{\dualf}{\!\!\!\!\left.\right.^\ast\!\! F}
\newcommand{\cdt}{\frac{1}{\dv}\pt}
\newcommand{\cdr}{\frac{1}{\dv}\pr}
\newcommand{\cdth}{\frac{1}{\dv}\pth}
\newcommand{\cdk}{\frac{1}{\dv}\pk}
\newcommand{\cdj}{\frac{1}{\dv}\pj}
\newcommand{\rad}{\;r\! a\! d\;}
\newcommand{\half}{\frac{1}{2}}
%%%%%%%%%%%%%%%%%%%%%%%%%%%%
%%%%    Here

\title[
  Is there a hidden connection between massive neutron stars  and dark matter in cosmology?]
{Is there a hidden connection between massive neutron stars  and dark matter  in cosmology? {}}
{}
\author[Hujeirat,  A. A.]
       {Hujeirat, A.A. \thanks{E-mail:AHujeirat@uni-hd.de} \\
\\
%  \footnotemark[1]\thanks{ }\\
%$^{1}$
IWR, Universit\"at Heidelberg, 69120 Heidelberg, Germany \\
%$^{2}$
}
\date{Accepted  ...}

\pagerange{\pageref{firstpage}--\pageref{lastpage}} \pubyear{2002}

\maketitle

\label{firstpage}

\begin{abstract}
 Astronomical observations reveal a gap in the mass spectrum of relativistic objects: neither black holes nor neutron stars with  2 - 5 solar masses have ever been observed.

 In this article I proceed in presenting the scenario which discloses a possible hidden connection between massive neutron stars (MANSs), dark matter and dark energy in cosmology. Accordingly, when the curved spacetime embedding MANSs compresses the
nuclear matter to beyond a critical supranuclear density $n_{cr},$ mesons, generally transmitting the residual nuclear forces between neutrons, could gain
energy by frequently  interacting with a scalar field $\phi$ at the background. When the effective energy of mesons becomes comparable to the bag energy enclosing the quarks, the neutrons merge together and form a super-baryon (SB), whose interior  is made of incompressible gluon-quark superfluid.
It turns out that the process has a runaway-character: it enables the super-baryon to grow in mass and volume from inside-to-outside to finally metamorphose the entire object into a completely invisible dark gluon-quark object, practically indistinguishable from isolated stellar black holes. The inability of these objects to merge with other objects whilst agglomerating in clusters makes them excellent candidates both for black holes and for dark matter halos in cosmology.

\end{abstract}


\section{Introduction:  cosmology of massive neutron stars?}
             Unlike luminous stars, whose energies are generated through nuclear fusion, neutron stars emit the rest energy
             stored in their interiors from old evolutionary epochs. As in the case of luminous normal stars, the total energy emitted by neutron stars is proportional to  the their masses, implying therefore that massive neutron stars must be short-living objects also.\\
%%%%%%%%%%%%%%%%%%%%%%%%%%%%%%%%%%%%%%%%%%%%%%%%%%%%%%%%%%%%%%%%%%%%%%%%%%%%%%%%%
\begin{figure}%[htb]
\centering {%  \hspace*{-0.5cm}
\includegraphics*[angle=-0, width=7.15cm]{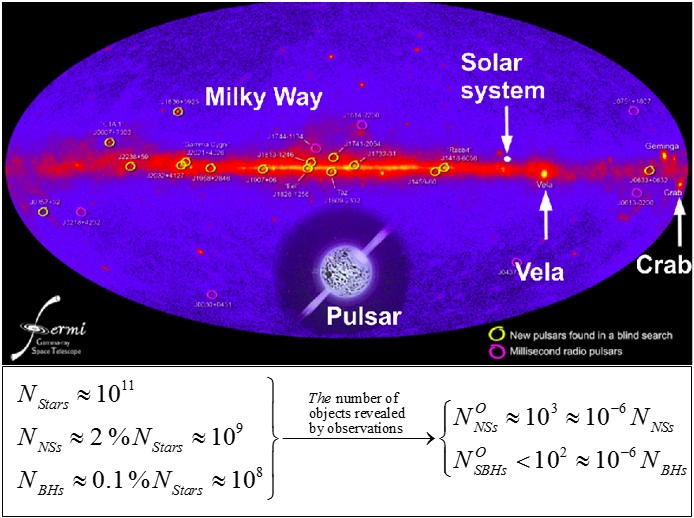}\\
}
\caption{\small  NSs and pulsars emit their radiation mainly in the X-ray and radio bands. Among the several billions of stars of the Milky Way,
only several thousands NSs and pulsars and about several hundreds stellar black holes  have been observed so far. These numbers are
about one million times smaller than those expected from theoretical and statistical considerations. }  \label{NS_MilkyWay}
\end{figure}
%%%%%%%%%%%%%%%%%%%%%%%%%%%
              Similar to the natural selection scenario of primates, most massive astrophysical objects must have disappeared relatively quickly, but only solar-like objects  are able to shine for billion of years and to be observable until the present universe:
              thanks to the parameters characterizing our universe.\\
              Just for illustration:  a ten solar masses star has  a lifetime 1000 shorter than that of the Sun.
              On the other hand, cosmological simulations reveal that the first stars must have been 100 to 10000 solar masses and
              that they should have formed from primordial clouds made solely of hydrogen \cite[][]{Bromm2004}. In the absence of heavy elements, it is believed
              that these massive stars must have collapsed directly into stellar black holes, but whose masses have been growing continuously
               through accretion of matter from their surroundings and/or through repeated mergers with other objects  to become the monstrous black holes that reside the centers  of almost all observable galaxies.\\
              However, an evolutionary track in which the first stars, or at least a part of them, may have collapsed to form pulsars and/or neutron stars
              statistically cannot be  excluded. Moreover, if the parameters characterizing our universe indeed do not allow matter-density to grow indefinitely
              \cite[][]{Nassif2016},
              then the abundance of massive neuron stars at that epoch must have been rich. Under these circumstances, the first
              generation of NSs must have emitted their energies long time ago to became invisible and disappear from our today  observational windows.\\
              Indeed, the following list of arguments are only a few  in favor of this scenario:\\

              \begin{itemize}
                \item Relativistic compact objects with $2~M_{\MSun}<M<6M_{\MSun} $ practically do not exist
                \item The number of  relativistic compact objects so far found to populate the Milky Way is approximately
                         one-million time smaller than expected from theoretical and statistical considerations \cite[Fig. 1,][]{Witten1984}
                \item The mass range of black holes is practically unlimited with neither lower nor upper bounds are known, whereas NSs enjoy an unusually narrow mass range.
                \item Isolated neutron stars that are older than one Gyr haven't been observed yet.
                \item Modelling the internal structure of NSs requires their central densities  to be far  beyond the nuclear density: an unknown density
                  regime  in which  most EOSs become physically inconsistent \cite[][]{Camenzind2007, Hempel2011}.
                \item All EOSs break down when  nuclear fluid becomes weakly compressible.
                \item The glitch phenomena observed in NSs and pulsars indicate that NS-cores are governed by superfluids \cite[Fig. 2,][]{Shapiro1983,Espinoza2011}.
              \end{itemize}
 %%%%%%%%%%%%%%%%%%%%%%%%%%%%%%%%%%%%%%%%%%%%%%%%%%%%%%%%%%%%%%%%%%%%%%%%%%%%%%%%%
\begin{figure}%[htb]
\centering {%  \hspace*{-0.5cm}
\includegraphics*[angle=-0, width=7.5cm]{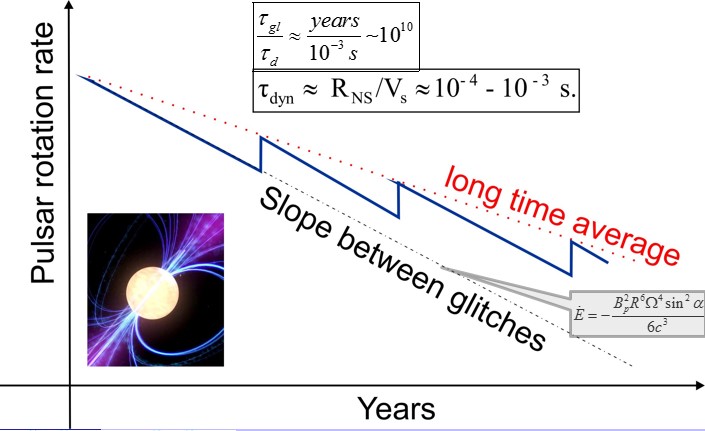}\\
}
\caption{\small NSs and pulsars have relatively strong magnetic fields that are intimately coupled to the rotation energy of these objects. As
magnetic diploe radiation is emitted, their rotational frequencies decrease and therefore they spin down continuously.
What is observed however,  is that these objects undergo repeated spin up that occurs abruptly almost every one, two or three years: a phenomena
 called pulsar glitches (Shapiro \&Teukolsky 1983).
Recalling that the sound crossing time through these objects is of order milliseconds, then it was a mystery  for experts to understand:
what keeps the object passive for almost 10 billion times the dynamical time scale, and then suddenly reacts. One of the proposed explanations was
that the nuclear matter inside the cores of these objects must be in a superfluid state and therefore are weakly coupled to the normal and dissipative crust.
 However, from time to time, the core must deposit certain amount of rotational energy into the crust, thereby causing a prompt speed up. }  \label{GlitchPhenomena}
\end{figure}
%%%%%%%%%%%%%%%%%%%%%%%%%%%
              As a consequence,  we expect isolated massive NSs to metamorphose into dark objects, whose interior are made of
              incompressible gluon-quark superfluids and to subsequently disappear from our observation windows.
\section{Normal dissipative fluids versus superfluids}
             Normal matter is usually made of self-interacting particles, non-ideal and dissipative medium. The illustrate exemplify these concepts, consider the flowing water in a river. Particles at the surface communicate with the motionless ones at the ground, generating thereby a velocity profile that varies with the depth, i.e. normal to the direction of motion. If we were to replace the water by honey, the profile of depth-depending velocity would change dramatically.
             The same
             applies for other materials, as each material has its own chemical and physical properties that determine the way particles communicate
              with each other. The collective effect is called friction, which mathematically  represented by anisotropic stress-tensor. The components normal to
              the direction of motions is called tension with the dynamical viscosity serves as a coefficient, inside which the chemical properties are encapsulated.\\
              The effect of viscosity is generally to speed up and/or slow down the motions of particles in different portions of the domain toward enforcing a uniform motion. But if the flow is subject to external (non-conservative) forces and the viscosity is sufficiently small, then the motion of the particles become random, where the entropy of the system saturates.  Such a fluid flow is said to be dissipative and therefore irreversible.\\
           %%%%%%%%%%%%%%%%%%%%%%%%%%%%%%%%%%%%%%%%%%%%%%%%%%%%%%%%%%%%%%%%%%%%%%%%%%%%%%%%%
\begin{figure}%[htb]
\centering {%  \hspace*{-0.5cm}
\includegraphics*[angle=-0, width=5.5cm]{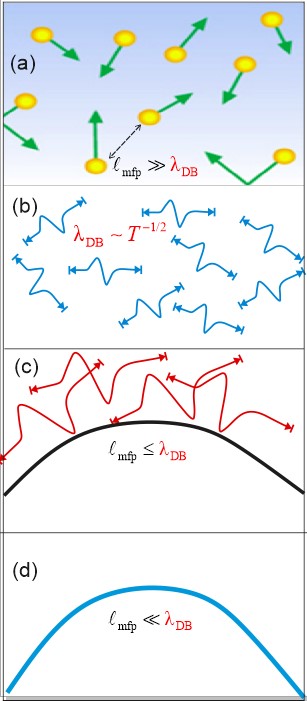}\\
}
\caption{\small  The rate and form of interactions between  particles  depend strongly on the temperature. According to quantum dynamics, each particle
 is associated with the De Broglie thermal wavelength, $\lambda_{DB}$, which is inversely proportion to the temperature. It turns out that at sufficiently low temperature, when $\lambda_{DB}$  becomes larger than the mean free path between particles, then they start sharing their quantum properties
with their neighbours.} \label{DeBroglie}
\end{figure}
%%%%%%%%%%%%%%%%%%%%%%%%%%%
              It turns out that when the temperature of the fluid falls below a certain critical value, the effect of viscosity diminishes. In this case, the fluid enters               the so-called superfluid phase, where quantum mechanical effects start to emerge on global scales, for example, climbing up the walls of the container or forming discrete number of vortices  that rotate coherently with each other.\\
              In such fluids the De Broglie wavelengths  $\lambda_{DB},$ surpasses the mean free path characterizing the collisions between particles, and then each particle starts to coordinate its motion with its neighbours to finally clothe their quantum state: an extraordinary phenomenon  in which micro-quantum states start showing up on the macroscopic scales (Fig. 3).\\
              In terrestrial fluids superfluidity phases start to show up  when the temperature of the fluid becomes
              approximately one-hundred times smaller than the corresponding Fermi-temperature. In the cores of neutron stars however, although the temperature is of order one hundred million degrees, nuclear fluids are still about ten thousand times lower than the corresponding Fermi-temperature, implying therefore that NS-cores most likely are in quantum superfluid phase.
              %%%%%%%%%%%%%%%%%%%%%%%%%%%%%%%%%%%%%%%%%%%%%%
\subsection{Astrophysics of weakly compressible and incompressible fluid flows}
               In an ever expanding universe the ultimate phase of nuclear fluids inside the cores of isolated NSs should have vanishing  entropy
               inviscid and incompressible.
               In the early evolutionary phases of NSs, their cores should be  threaded by vortex lines, where the rotational energy is stored. In this case,  Kelvin waves
               in combination with buoyancy effects are expected to be the dominant  transporter of energy from their interiors into the surrounding media.
               In fact, computer simulations of rotating superfluids reveal that the enclosed vortices are not static, but oscillate and intersect with each other
               to finally turn the configuration turbulent \cite[Fig. 4, ][]{Baranghi2008,Baggaley2014}.
               Based on these observations, we don't expect the 10-kilometer long vortex lines threading the cores of NSs and pulsars to behave differently.
               As a consequence,  turbulent motion of vortex lines in superfluids generally enhances dissipation of rotational energy
               to subsequently lower their energy state.\\

%%%%%%%%%%%%%%%%%%%%%%%%%%%%%%%%%%%%%%%%%%%%%%%%%%%%%%%%%%%%%%%%%%%%%%%%%%%%%%%%%
\begin{figure}%[htb]
\centering {%  \hspace*{-0.5cm}
\includegraphics*[angle=-0, width=7.5cm]{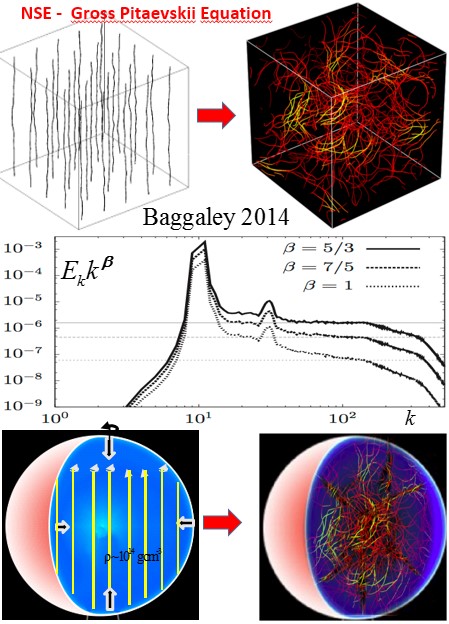}\\
}
\caption{\small Numerical modelings of vortex dynamics  in terrestrial superfluids reveal that vortices become turbulently unstable
and  their developed spectrum resembles that observed in normal fluids, though cascading proceeds at a lower rate. Thus the expected
trillions of kilometer-long vortices  threading NS-cores are unlikely to behave differently. This would speed up loss of rotational energy
of massive NSs and let them disappear from our observational widows on times scale
comparable to 1Gyr or even shorter.}\label{SuperfluidTurbulence}
\end{figure}
%%%%%%%%%%%%%%%%%%%%%%%%%%%
              Still, we have still to determine whether nuclear fluids inside the very central regions of NSs are \textbf{compressible or incompressible}
              \cite[see][and the references therin]{Hujeirat2009}.
              I should note that if the fluids inside the cores of these objects were compressible and stratified, then energy may be still extracted via acoustic waves  from their interior  into the crust or the surrounding media, thereby weakening the compressibility of the fluid even more.
              In fact such an energy extraction process is considered to be the underlying heating mechanism  of the solar corona \cite[][]{Bingham2010}.

               In the case of NSs, I  argue that incompressibility is an inevitable  phase of matter once the number density becomes larger than the nuclear one.
               Among the reasonable arguments that favor this phase are the following:\\

               \begin{enumerate}
                 \item  The  spatial variation of  the coefficient $g_{rr}$ of the Schwarzschild metric on the length scales of atomic nuclei is roughly
                  ($dg_{rr}/dl \ll 10^{-19}$) implying therefore that gravity-induced stratification is  unmeasurably small.
                                              %%%%%%%%%%%%%%%%%%%%%%%%%%%%%%%%%%
                 \item The effective potential of the gluon-field inside individual baryons is predicted to increases with radius as $r^{\Gamma(\geq 1)}$.
                 Thus the gluon-quark effective force inside hadrons opposes compression by gravity
                                               %%%%%%%%%%%%%%%%%%%%%%%%%%%%%%%%%%
                 \item   Most EOSs used for modeling the very central regions of NS-cores display sound velocities that do not respect causality.
                   However, fluids with $V_s = \calO(c)$ cannot be compressed anymore. In fact recent numerical simulations of classical incompressible Navier-Stokes fluid-like  flows reveal a blatant inconsistency in the capturing flow configurations, whenever  the employed EOSs
                   are set to depend on the local properties of the fluid only \cite[][]{Bechtel2004,SAEAIR1990}.
                          In this case communicators that merely depends on local exchange of information are insufficient for efficiently coupling  different/remote parts    of  the fluid in a physically consistent manner.
                       A relevant example is the solution of the TOV-equation for the incompressible case, where  the internal energy density, $\calE $ is set
                       to be  constant.   The pressure here turns out to depend on the global compactness of the object, but it becomes  even acausal when the global compactness of the object is enhanced \cite[][]{Glendenning2007}.\\
                      Thus, using a local description of the pressure for simulating  weakly compressible or incompressible fluids is physically  inconsistent.
                                           %%%%%%%%%%%%%%%%%%%%%%%%%%%%%%%%%%
                 \item Beyond the nuclear density,  most sophisticated EOSs tend to converge to the limiting EOS: $P_L = \calE$ (Fig. 5), where $P_L, \calE$ denote the local pressure and the enery density, respectively. However, such  fluids
                   cannot accept compressibility anymore as otherwise the causality condition would be violated.  In this case  the nuclear fluid must obey the
                   EOS: $\calE = a~n^2,$ where n is the number density. When taking the regularity condition of the pressure at the center of the object into account, i.e.,
                   $\nabla P |_{r=0}=0, $ one finds that there is a maximum critical number density $n_{cr},$ at which both the Gibbs function as well as its derivative
                   vanish.
                   In a previous work \cite[][]{Hujeirat2016}, it was shown that $n_{cr}= 3 n_{0}$, where $ n_{0}$ is the nuclear density. For $n \geq n_{cr} $ the nuclear fluid becomes purely incompressible. Such fluids are expected to form, when all other forms of energies, e.g. kinetic$(E_{kin})$, magnetic $(E_{mag})$ and thermal $(E_{th})$ energies have been evacuated out of the very central region of  NSs.
                   One may think of these energies as perturbations superimposed on a zero entropy state at the background.

%%%%%%%%%%%%%%%%%%%%%%%%%%%%
\begin{figure}%[htb]
\centering {%  \hspace*{-0.5cm}
\includegraphics*[angle=-0, width=6.5cm]{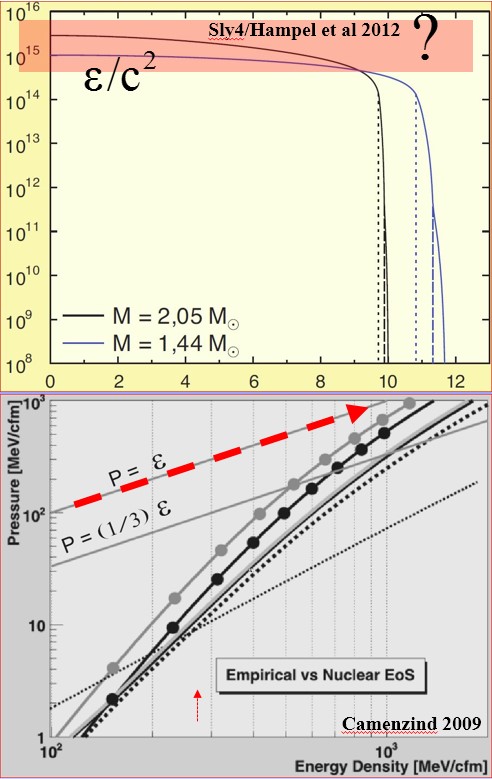}\\
}
\caption{\small  According to numerical and theoretical studies,  the bulk of matter in  NSs
must have densities beyond the nuclear one, though the physical properties of matter in this density regime are poorly understood.
 Nevertheless, in this supranuclear density regime most  EOSs appear to converge  towards the stiffest possible EOS: $P = \calE = a_0 n^2.$
 Fluids governed by such a critical EOS  become purely incompressible.
    }  \label{EOSs_Convergence}
\end{figure}
%%%%%%%%%%%%%%%%%%%%%%%%%%%

                   The classical form of the first law of thermodynamics: $ dE= Tds - pdVol $  is not valid for ncompressible nuclear fluids   as both $dE$  and
                   $dVol $ are unrelated  and therefore the local pressure $P_L$ cannot be calculated from  $dE/dVol.$ \\
                   In fact the imposed regularity condition on the pressure  at r=0 manifests the incompressibility character of the  fluid and therefore the break down of  formula  $dE/dVol = -P_L= -n^2 \DD{\D}{\D n}(\calE/n).$
                   Moreover, assuming the matter at the center to obey the EOS: $P=\calE,$ then the  TOV equation can be integrated to yield:  $\calE(r) e^{\calV(r)} = const.,$ where $\calV$ stands for the gravitational potential.
                   However,  nuclear matter obeying $P=\calE,$ cannot be compressed and therefore ${\calV}$ must be constant, which means a vanishing gravitation-induced stratification. Nevertheless, most NS-models rely on using a non-vanishing density-gradient even at the vicinity of $r=0.$
                   In order to have a NS of a reasonable mass, the central density must be much beyond the nuclear density:
                   a density regime which is experimentally untestable and where our theoretical knowledge is severely limited (Fig. 5).
                 Consequently, the existence of a non-local pressure in an environment, where the fluid is weakly compressible , such as in the vicinity
                  of  the center of NSs is necessary in order to escape their collapse into  black holes.\\
               \end{enumerate}
%%%%%%%%%%%%%%%%%%%%%%%%%%%%
\begin{figure}%[htb]
\centering {%  \hspace*{-0.5cm}
\includegraphics*[angle=-0, width=5.5cm]{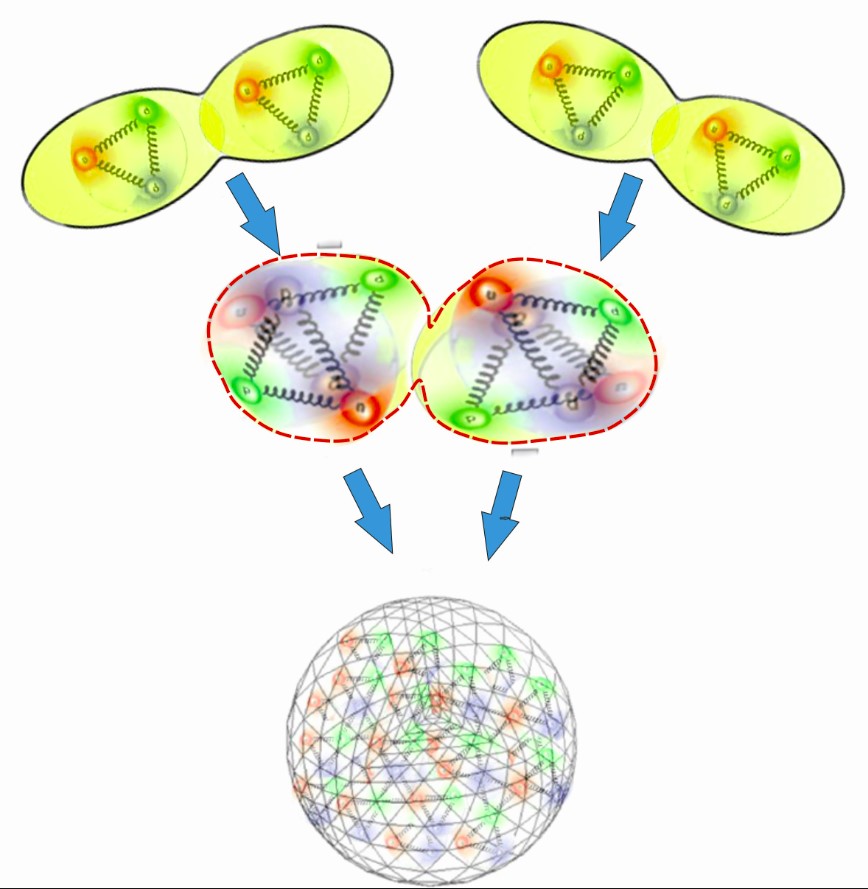}\\
}
\caption{\small When two neutrons  are brought together to merge, the employed force should be equal or even larger than the energy required for their creation. The employed effective energy is then absorbed and used to form new communication channels between quarks, thereby enhancing the effective mass of the newly formed super-baryon. This is in line with experimental data which revealed that the effective energy of the short-living pentaquarks correlates almost linearly with  the number of communication channels, through which the strong force is communicated.}  \label{BaryonMerger}
\end{figure}
%%%%%%%%%%%%%%%%%%%%%%%%%%%
%%%%%%%%%%%%%%%%%%%%%%%%%%%%%%%%%%%%%%%%%%%%%%%%%%%%%%%%%%%%%%%%%%%%%%%%%%%%%%%%%
\begin{figure}%[htb]
\centering {%  \hspace*{-0.5cm}
\includegraphics*[angle=-0, width=7.15cm]{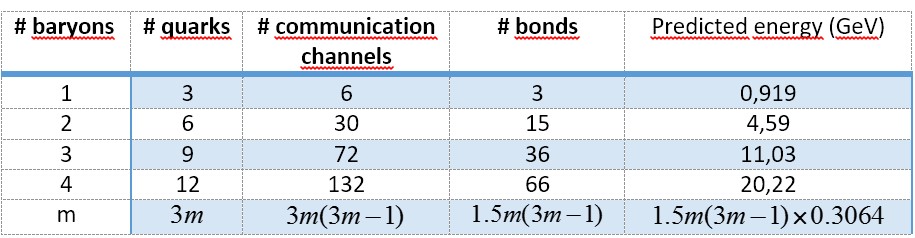}\\
}
\caption{\small  When baryons get to merge, then the number of communication channels,  through which the strong force  between
quarks is communicated, increases dramatically.   The newly created flux tubes  increase the  effective mass of the resulting super-baryon,
whose total energy should be larger than the sum of energies of the participating baryon, but still upper-bounded by  factor two.
}  \label{CommunicationChannels}
\end{figure}
%%%%%%%%%%%%%%%%%%%%%%%%%%%
 In fact, it appears that nuclear fluids with $n_{cr}\geq 3 n_{0}$,  having  a constant internal energy density and $E_{kin}=E_{th}=E_{mg}=0$
 should be incompressible gluon-quark superfluids  \cite[see][and the references therein]{Hujeirat2016}. Under such conditions all sorts of cooling, that
 contribute to entropy generation, including neutrino emission
via Urca processes, will be suppressed.\\
 Indeed, recent experimental observations from the Relativistic Heavy Ion Collider (RHIC) and the Large Hadronic Collider (LHC) revealed that when heavy ions are accelerated in opposite directions to reach roughly the speed of light and collide, the outcome was not a gas, but
 rather a nearly frictionless superfluid, even though the effective temperature was extraordinary high
 \cite[see][and the references therein]{Nagle2011}. Therefore, zero-temperature quantum fluids with number densities $n_{cr}\geq 3 n_0$  are expected to be incompressible superfluids.\\
 \section{But what makes DEQOs disappear from our observational windows and how can they still escape collapse into BHs?}
  The brief answer is that when the number density of zero-temperature quantum fluids surpasses the  critical density $n \geq n_{cr},$ the sub-nuclear particles, such as mesons and gluons start interacting with the scalar field more frequent, thereby enhancing their effective mass and
  enabling individual baryons to merge together to form  a super-baryon. The potential governing its interior increases with radius, giving rise
  to   a non-local negative pressure that apposes compression. Consequently, the compactness  of the object is significantly enhanced  and therefore
  the object sinks deeply into the embedding curved spacetime to finally disappear from all direct observational windows.\\

\textbf{\emph{\emph{What is the underlying physical mechanism for generating "dark energy"  in NSs? }}}\\

The contribution of quarks to the baryon mass is approximately $2\%,$  whereas the energy required to deconfine them is roughly equal or even larger than
  the energy needed  for the creation of the whole baryon, which is roughly equal to  $0.94\,GeV.$\\
  However, as gluons are virtual particles that are generated by vacuum fluctuations that popping into existence and disappearing,
 then the symmetry between creation and annulation must be perfectly tuned, as otherwise protons would not survive a life time
 of the order $10^{52}$ the light crossing time through a baryon. Thus the numerous complicated interactions between the sub-nuclear particles embedded
 in the  quark-gluon cloud  are perfectly organized and fine-tuned, so that energy loss via dissipation is completely suppressed. This implies that
 quark-gluon plasma acquires a one single quantum state. In this case, the entropy $ dS= k_B log \Omega$ must vanish, where $\Omega$ is the number of all possible microscopic states. This is in line with experimental data of the  RHIC and the LHC, which
 showed that slamming heavy ions against each other with almost the speed of light produced almost a frictionless fluid with very low entropy.\\
 In fact,  particles of a plasma that rely on different particle-mediators with various communication speeds, e.g. the speed of light,
  speed of sound $V_s << c$, transport velocity, viscous interaction speed and so on, then their direct and indirect collisions generally lead to random motions  of remote fluid parcels that subsequently
  turn turbulent, where entropy generation is enhanced. This agrees well with
 numerical simulations,  which show that high Reynolds number flows may turn turbulent, whenever the internal interactions between fluid parcels are
 mediated with different communication speeds\cite[][]{Kerstin2011}. The cascading  mechanism by which turbulent kinetic energy is transferred into other
energy forms,  will still increase  entropy. Consequently, in order keep baryons stable and suppressing entropy generation, the interactions between the constituents of quark-gluon plasmas  should be fine-tuned and maintained  via one single speed, namely the speed of light.

As the nuclear fluid in the very central region of NSs has density beyond the nuclear one, the only degree of freedom left, where exotic  energy
could be still created would be through merging baryons and generating new communication channels between the quarks. This however requires
enhancing compression by external forces, e.g. enhancing the curvature of the embedding spacetime (Fig. 12). Similar to the recently explored pentaquarks,
new communication channels must be generated between the quarks to ensure stability of the internal structure of the newly born super-baryon. The energy required for
constructing the channels comes mainly from quark-antiquark interactions with the field, which we term here as dark energy.\\
%%%%%%%%%%%%%%%%%%%%%%%%%%%
When the first baryons at r=0  merge with its neighbors and form a new super-baryon, the number of communication's channels, i.e. the flux tubes through which the strong force is communicated between the quarks, increases non-linearly with the number of participating quarks. For instance, for a given number of baryons, say $n,$ each which is made of
 three quarks flavor, the number of communication channels scales as  $n(n-1)\sim n^2,$ so that the number of bonds between the quarks increases as
 ${n(n-1)}/{2}$  (Fig. 7).  As the bonds between quarks are the ones that contribute mostly  to effective mass of the super-baryons, rather than the quarks themselves, the process is equivalent to injection of dark energy.  The excess of dark energy goes to increase  the surface tension as well as
  the volume energy of the bag enclosing the freely moving quarks.  Hence the total work , $dW_{tot}$ per unit volume $dVol$ needed to increase the volume of  the super-baryon  reads:
 \beq
 \DD{dW_{tot}}{dVol} = -P +    \vec{n}\cdot \textbf{dT} ,
 \eeq
where $P,~\textbf{dT}$ denote the local pressure and the surface tension perpendicular to the normal vector $ \vec{n}$ generated by the enclosed quarks $dN,$  respectively.
Taking into account that the concerned super-baryon is spherically symmetric and composed of incompressible fluids ($P,~dN/dVol = const.$),
 we obtain that $\vec{n}\cdot \textbf{dT}= \alpha_\Phi  r^2$  and therefore the dark energy density would have the form:
  \beq
           \calE_\Phi = \alpha_\Phi r^2 + \beta_\Phi,
 \eeq
 $\alpha_\Phi, \beta_\Phi $  are  constants.
 In fact this is  similar to the static quark-antiquark  potential inside individual baryons, which can be described as a superposition of  Coulomb-like term, i.e, $(\, const./r)$  and  a term, whose action increases with radius. \\
 As the concerned quantum fluid is incompressible, in hydrostatic equilibrium and in a superfluid phase, the scalar field, which is the source of dark energy, can be safely considered as a spatially and temporally constant. This implies that the EOS of dark energy is $P_\Phi = -\calE_\Phi,$ i.e. extremely stiff, non-local and behaves like $ r^2.$
As the mass of the super-baryon continues to grow via merger with the surrounding baryons as well as through the injection of dark energy, the negative pressure must increase as $r^2,$ to finally attain a global maximum at its surface of the object.

 On the other hand, astronomical observations reveal that the compactness parameter
 of most NSs and pulsars are equal or even larger than half. This however imposes a constrain on the EOS of dark energy in NSs:  a non-local negative pressure of the form $P_\Phi = - \alpha \calE_\Phi,$ with $\alpha<1$  should be excluded, as otherwise these objects would collapse into solar-mass BHs,
 whose existence is not supported by observations.

\section{Determining the parameter regimes: the connection between quantum and GR scales}

As we have mentioned already, gluon-quark superfluid inside baryons must be incompressible. It is however not clear
how spatially separated superfluid parcels could be brought together to merge without violating the incompressibility character of the fluid?
One could envisage an instantaneous crossover phase transition in which the compressible nuclear fluid consisting of individual baryons and
 having chemical potential $\mu \sim n $ turns into incompressible gluon-quark-superfluid with $\mu \sim const. $  (Fig. 8). Here the injection of dark energy
 plays the role of a catalyst, i.e. the  instantaneous change of the EOS  must run as follows:
 \beq
  \calE = a_0 n^2  \xrightarrow{\textrm{dark energy}}   \calE = a_{qsf} \times n,
 \eeq
 where $a,~a_{qsf}$ are constant coefficients. Note that the injection of dark energy is necessary for boosting the energy of mesons and to subsequently convert them into gluons needed for forming the new flux tubes between the quarks inside the super-baryon.

%%%%%%%%%%%%%%%%%%%%%%%%%%%%%%%%%%%%%%%%%%%%%%%%%%%%%%%%%%%%%%%%%%%%%%%%%%%%%%%%%
\begin{figure}%[htb]
\centering {%  \hspace*{-0.5cm}
\includegraphics*[angle=-0, width=6.5cm]{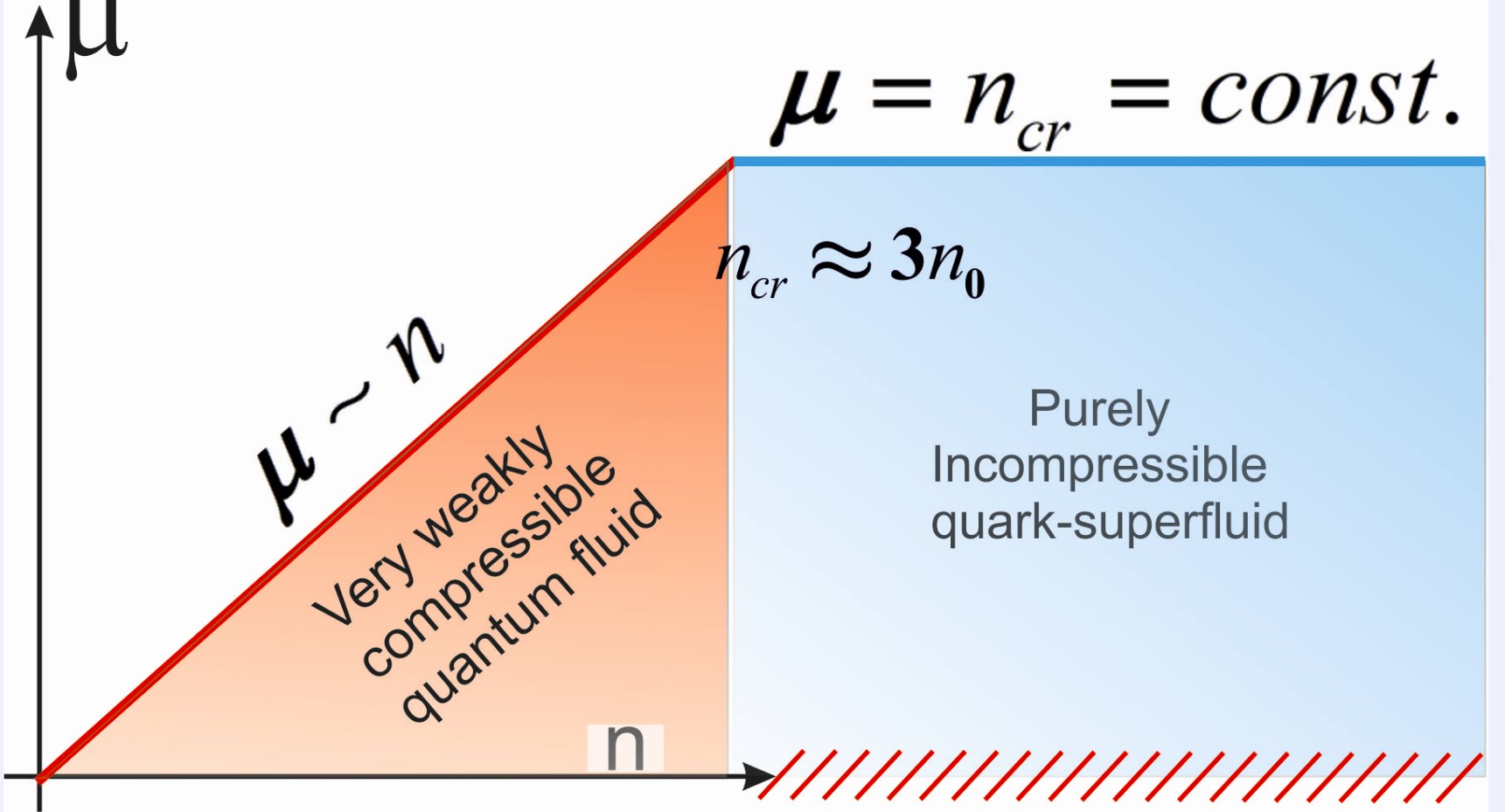}\\
}
\caption{\small  At the very center of NSs and pulsars the number density is predicted to be beyond the nuclear one $n_0.$ On the other hand,
 most EOSs in this density regime appear to converge asymptotically to $ \calE \sim n^2 ,$ where the nuclear fluid becomes incompressible as the
 sound speed reaches the speed of light.
In this case the chemical potential increases linearly with the number density to finally reaches the upper bound
critical density $n_{cr}= 3\,n_0,$ beyond which the chemical potential remains constant.  This corresponds to the state in which the normal baryonic particles merge together to form a single super-baryon, whose interior is made of incompressible gluon-quark superfluid.
  }  \label{FaceTransition}
\end{figure}
%%%%%%%%%%%%%%%%%%%%%%%%%%%

In order to insure that the dark energy goes to solely enable a smooth crossover phase transition, we require that there must be a
critical number density $n_{cr},$ where the Gibbs function vanishes.  The combined energy density
 (i.e., the density of internal energy of  baryons and that of dark energy $\calE_\phi$ )
per particle should be larger than or equal to the energy required to de-confine the quarks inside individual baryons:
  \beq
                                f(n) = \DD{\calE_b + \calE_\phi}{n} - 0.939 ~GeV \geq 0.
  \eeq
  Using the scalings $[\rho]=10^{15}g/cm^3~(\doteq 0.597/fm^3),$ chemical potential (energy per particle) $[\mu] = 1\,GeV,$  we then obtain
  $[a_0]=1.674~GeV\, fm^3$ and  $ [a_\phi]= 5.97\times 10^{-39}~GeV/fm^5$ and $[b_\phi] = 0.597~GeV/fm^3.$\\
  In this case the  Gibbs function in non-dimensional units reads:
\beq
                                f(n) = a_0 n + \DD{ b_\phi}{n} - 0.939.
  \eeq
The function $f(n)$ may have several minima, depending on the values of $a_0$ and $b_\phi.$  However, for a crossover phase transition to occur, both
$f(n)$ and $\D f(n)/\D n $ must vanish, which occurs at  $n=0.81$  for the most reasonable values:  $a_0=1$ and $b_0 = 0.37$ (Fig. 9).\\
Thus the potential of vacuum energy at r=0 is  $ V_\phi = b_\phi,$ and therefore there is a  non-local pressure $P_{NL}= -\calE_\phi= -V_\phi = - b_\phi.$
For determining the value of $a_\phi,$  we need to study the ultimate global structure of  the object.\\
             %%%%%%%%%%%%%%%%%%%%%%%%%%%%%%%%%%%%%%%%%%%%%%%%%%%%%%%%%%%%%%%%%%%%%%%%%%%%%%%%%
\begin{figure}%[htb]
\centering {%  \hspace*{-0.5cm}
\includegraphics*[angle=-0, width=6.95cm]{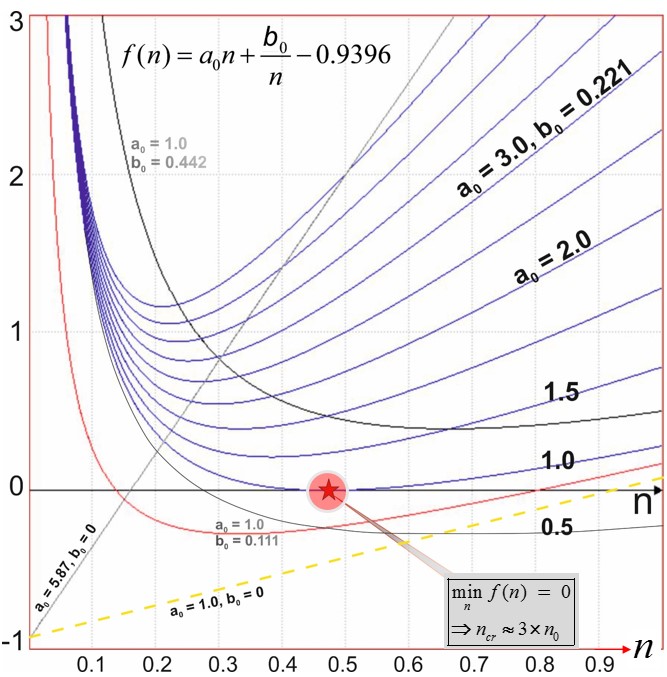}\\
}
\caption{\small$\calE = a_0\,n^2$  corresponds to the limiting EOS in the supranuclear density regime, where baryons may merge together
to form super-baryons.  The injection of dark energy goes mainly to create communication channels
connecting the quarks. The global effect of the injected dark energy takes the form of a non-local negative pressure that
apposes compression by the embedding spacetime. When analyzing the Gibbs function, we find that it attains zero-minimum at roughly three
times the nuclear density, i.e. $n_{cr}\approx 3\,n_0.$ Note that  for a crossover phase transition to occur, both the Gibbs function and its derivative
must vanish. }  \label{CriticalNumnerDensity}
\end{figure}
%%%%%%%%%%%%%%%%%%%%%%%%%%%

%%%%%%%%%%%%%%%%%%%%%%%%%%%%%%%%%%%%%%%%%%%%%%%%%%%%%%%%%%%%%%%%%%%%%%%%%%%%%%%%%
\begin{figure}%[htb]
\centering {%  \hspace*{-0.5cm}
\includegraphics*[angle=-0, width=7.5cm]{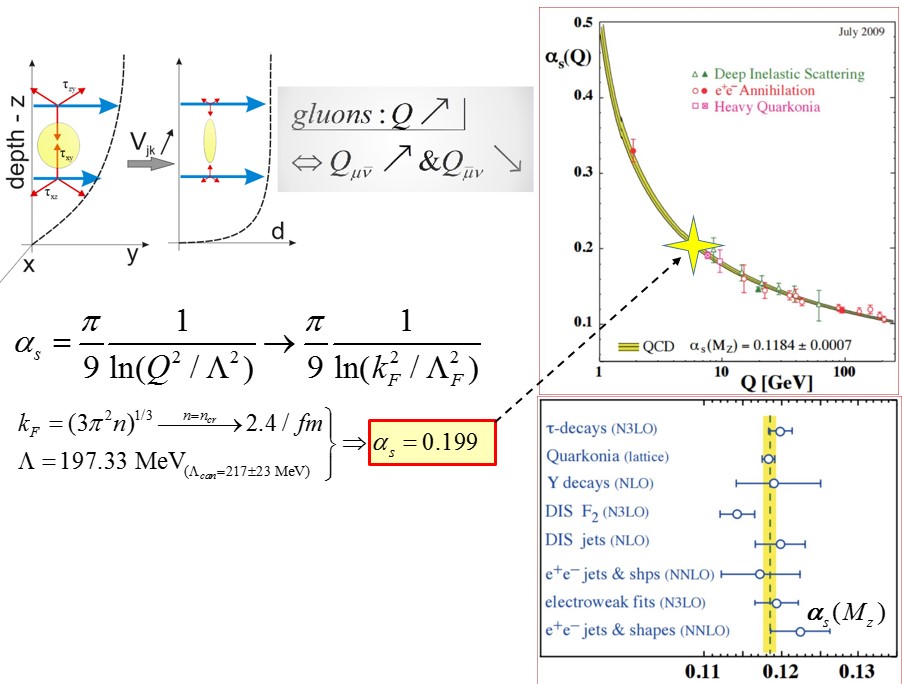}\\
}
\caption{\small  The matter density inside DEQOs  correspond to three times the nuclear density. While experimental data predict the coupling constant $\alpha_{QCD}$ to be around 0.12,
in the present case we find that $\alpha_{QCD}$ should be around 0.199. We attribute the difference to the non-overlapping density regimes
 used in both approaches, to the effect of dark energy  and to the global compression by the surrounding curved spacetime.   The effect of the coupling constant in QCD may be viewed as a measure for  reducing
 the perturbative velocity components  of quarks normal to their direction of motions. At sufficiently high energy/momenta, particles prefer
 to asymptotically move in parallel, where entropy generation diminishes and the gluon-quark plasma turns into an incompressible
 superfluid. }  \label{AsympFreedom}
\end{figure}
%%%%%%%%%%%%%%%%%%%%%%%%%%%
We note that the radius of the super-baryon behaves like a transition front that propagates outwards through the ultra-weakly compressible nuclear fluid of the NS, leaving the matter behind its front in an incompressible gluon-quark-superfluid phase. When the front reaches the
surface of the entire object, which is expected to occur  on the scale of $ \calO (10^8)\,yrs,$  then the object becomes a DEQO and disappears from our observational windows, as its radius would be  indistinguishable from the corresponding event horizon. Equivalently, we require the following equation to be fulfilled:

\beq
R_{*q} = R_S =  \DD{2 G_g}{c^2}(\calM_{NS}+ \calM_\phi),
\eeq
where $R_{*q}, ~R_S,~\calM_{NS} \textrm{~ and ~}\calM_\phi$ denote the radius of the object, Schwarzschild radius, Mass of the original NS
and the mass-enhancement due to dark energy, respectively. As the number density inside the DEQO is constant and equal to $3\times n_0$ and
as the vacuum energy density obey the relation $\calE_\phi = \alpha_\phi r^2 + \beta_\phi,$ then:
\beq
\barr{lcl}
\calM_{NS} &= & 4\pi \int_0^R \calE_b r^2 dr = \DD{4\pi}{3} \rho_{cr} R^3  \\
\calM_\phi  &= & 4\pi \int_0^R \calE_\phi r^2 dr = \DD{4\pi}{5} \alpha_\phi R^5 + \DD{4\pi}{3} \beta_\phi R^3.
\earr
\eeq

Let us nondimensionlize Eq. (6)  using the following scaling values:
$ [R] = 10^6 cm,\,  [\beta_\phi]=[\rho]=10^{15}\,g/cc, [\alpha_\phi]= [\rho/R^2],\tilde{\calM}=[\calM]= \DD{4\pi}{3}[\rho][R] ^3 = 2.1\MSun.$
We then obtain the following equivalent form to Eq. (6):
\beq
\DD{M_\phi}{M_{NS}} = \DD{3}{5} (\DD{R^2_{q\star}}{\rho_{cr}})\alpha_\phi +       \DD{\beta_\phi}{\rho_{cr}}.
\eeq

%%%%%%%%%%%%%%%%%%%%%%%%%%%%%%%%%%%%%%%%%%%%%%%%%%%%%%%%%%%%%%%%%%%%%%%%%%%%%%%%%
\begin{figure}%[htb]
\centering {%  \hspace*{-0.5cm}
\includegraphics*[angle=-0, width=7.5cm]{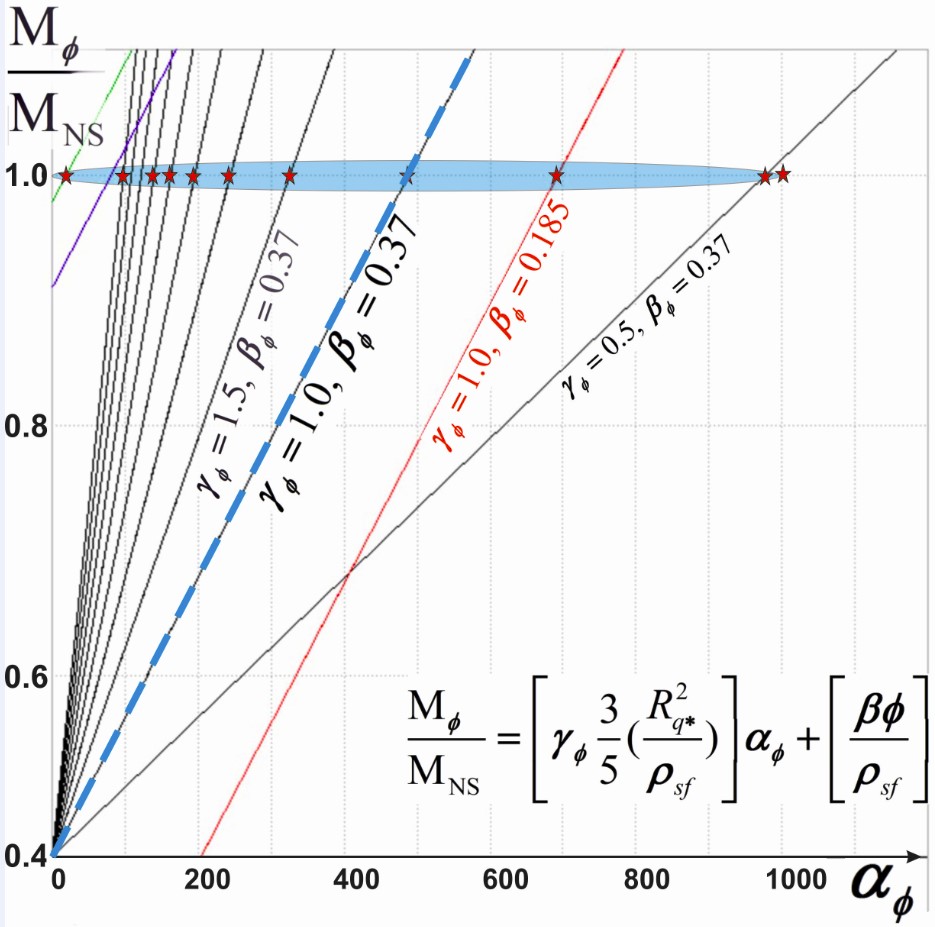}\\
}
\caption{\small  The ultimate total mass of the object relative to its original  baryonic mass versus the black energy coefficient $\alpha_\phi$ is shown   for  different values of the bag energy $\beta_\phi$ and manual enhanced weight $\gamma_\phi$ of $\alpha_\phi.$  According to QCD, the value of $\beta_\phi$ should lay around
  0.221 GeV, which is equivalent to 0.37 in the here-used non-dimensional units. The value: $\alpha_\phi = 490$ gives rise to an object whose radius
  coincides with the corresponding event horizon of the object.  This value of $\alpha_\phi$ is crucial for the dynamical stability of the object, as
  other low or high values  would eventually lead to its self-collapse into a BH. }  \label{MassVersusAPhi}
\end{figure}
%%%%%%%%%%%%%%%%%%%%%%%%%%%

 The term $\alpha_\phi\, r^2$ in Eq. (2) is the source of the  non-local vacuum pressure, which yields the incompressibility character of the gluon-quark superfluid in
 self-gravitating systems.\\
           Recalling that numerical and theoretical studies of the internal structure of NSs predict a compactness parameter
           $ \alpha_s (\doteq R_s/R_{NS}) \geq 1/2,$ which, in combination with the requirement that the object should turn invisible at the end of its cosmological life time, we conclude that its  final total mass $\calM_{tot} \leq 2 \times \calM_{NS}.$  As it is shown in Fig. (11), and displayed as
           blue dashed line,      $\alpha_\phi= 490$ appears to safely  fulfill these constrains.\\

To summarize the parameter determination procedure:
\begin{enumerate}
  \item Let the isolated NS has the mass $M_{NS}.$
  \item The baryonic fluid at the verge of phase transition obeys the EOS $\calE_b = a_0 n^2,$ whereas the EOS of the gluon-quark superfluid
  is  $\calE_b = const.$ and  $\calE_\phi = \alpha_\phi\, r^2 +  \beta_\phi$  for the dark energy.

  \item From the minimization requirement of  the Gibbs function
       we obtained  the coefficient $a_0$ and $\beta_\phi,$ where the latter was set to equalize the bag constant in terms of the  MIT-description
       of quarks in QCD. Here we use $B^{1/4}= 220 ~MeV,$ which is equivalent to 0.37 in the here-used non-dimensional units.
  \item The coefficient $\alpha_\phi$ has been determined by requiring that the radius of the original NSscoincides with the corresponding
  Schwarzschild radius after  its metamorphosis into a DEQO.

\end{enumerate}

   %%%%%%%%%%%%%%%%%%%%%%%%%%

\section{How could DEQOs be connected to dark matter and dark energy in cosmology?}
Baryon matter in QCD is made of  gluon-quark plasmas \cite[][]{Bethke2007}. The quarks themselves however make merely $2\%$ of the baryon mass, whereas the remaining $98\%$ are from the field and other related sources. Hence the flux tubes governed by gluons
are the ones that grant neutons most of their effective masses. Indeed, this is in line with experimental data from the LHC during the years 2009-12, which  reveals that pentaquarks have been detected in the range between 4.38 - 4.45 GeV \cite[][]{LHCb2015,Roca2015}. Obviously they are more massive than the sum of just two individual baryons. The increase of effective mass appears to correlate with the number of  communications channels of the gluons connecting the quarks.
Here, instead of just 12 in two distinct baryons, there are  30 channels in hexaquarks, i.e. 15 bonds.  Assuming the energy stored in each bond
connecting two arbitrary quarks to be  $(0.938/3)~ GeV/channel = const.$ then a super-baryon consisting of hexaquarks would have roughly the energy:
 $15  \times (0.938/3)~ GeV/channel\,\approx \,4,6\,GeV,$ which is  only slightly higher than the value revealed from pentaquark. \\
However, due to the strong confinement effect, quarks and gluons exist exclusively inside baryons and never in free space.\\

 %%%%%%%%%%%%%%%%%%%%%%%%%%%%%%%%%%%%%%%%%%%%%%%%%%%%%%%%%%%%%%%%%%%%%%%%%%%%%%%%%
\begin{figure}%[htb]
\centering {%  \hspace*{-0.5cm}
\includegraphics*[angle=-0, width=4.75cm]{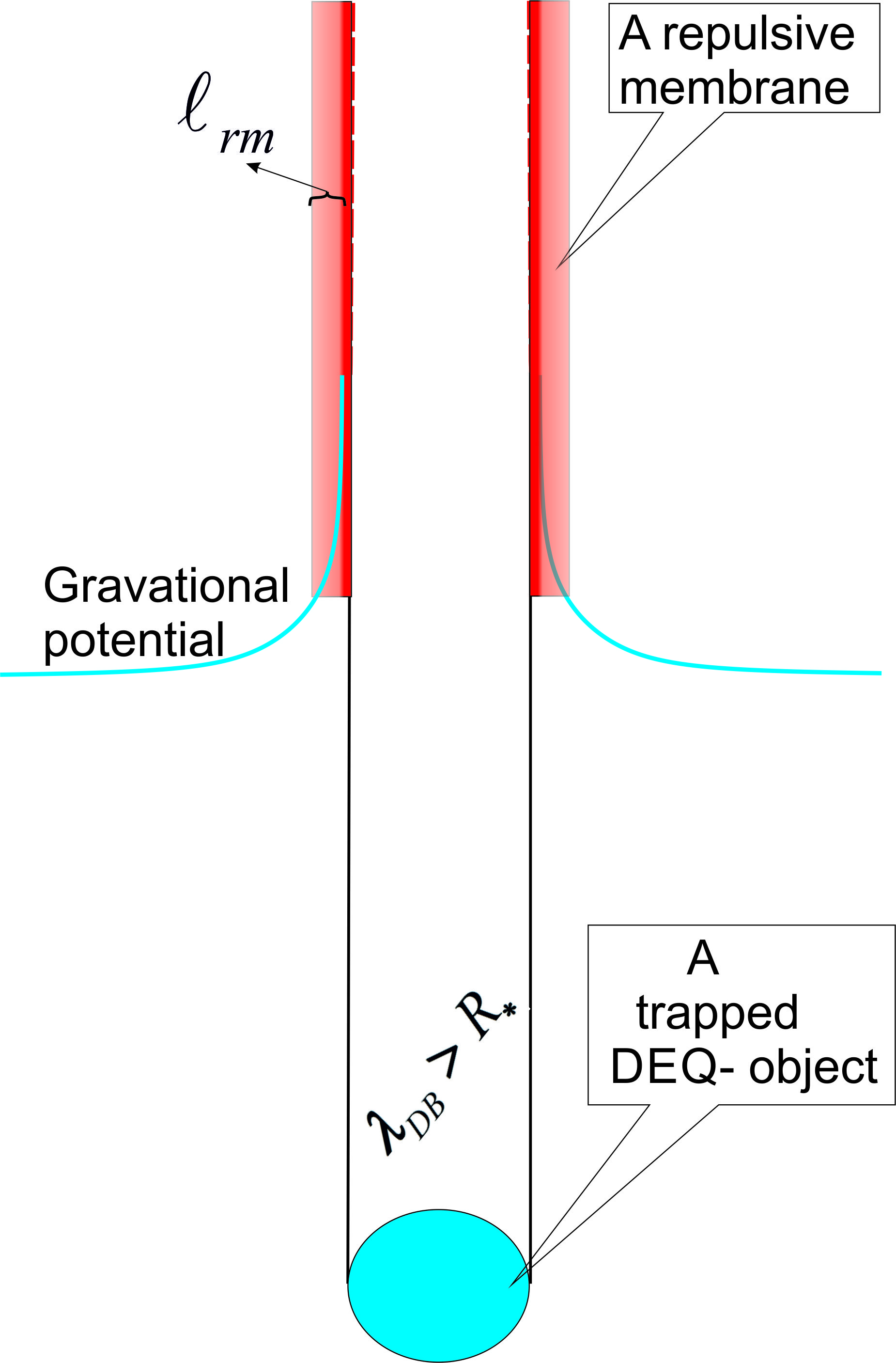}\\
}
\caption{\small   A DEQO made of incompressible gluon-Quark superfluid  deeply trapped in curved spacetime.
While the gravitational interaction enables the objects to agglomerate into clusters, at very small separation distances, i.e. $d = \calO( \Lambda_{rm}),$  the repulsive force dominates over other energies and inhibits direct interaction of  DECOs. The strength of the repulsive force  is proportional to the enclosed number of quarks, which roughly equals to their total deconfining energy.}  \label{TrappedGQSF}
\end{figure}
%%%%%%%%%%%%%%%%%%%%%%%%%%%
In the here-presented model, the density of matter at the very central region of  massive NSs is beyond the nuclear density, and therefore mergers of baryons
to form super-baryons cannot be excluded. \\
As more baryons are dissolved and join the super-baryon,  its volume and mass will increases to finally reach the surface of the entire object on the cosmological time scale.

Similar to gluon-quark plasmas inside individual baryons, the ocean of  the incompressible gluon-quark superfluid inside the
object would be shielded from the outside world by a repulsive quantum membrane, whose strength is proportional to the  number of the enclosed quarks
 (Fig. 12).\\
We conjecture that this membrane, which would be located at the horizon,would be sufficiently strong to prohibit quantum tunnelling
of particles both from inside and outside the wall, except for gravitons. If this is indeed the case, then there must be a length scales $\Lambda_{rm}$, so that when the separation length, $d,$ between two arbitrary DEQOs is comparable to $\Lambda_{rm},$
the objects would experience repulsive forces similar to those operating between individual baryons in atomic nuclei.

 Hydrodynamically, the generation of the dark energy inside the cores of massive NSs can be  modelled by introducing a scalar field, which,
 together with the baryonic energy, may be used to solve the TOV-equation inside these general relativistic objects \cite[][]{Hujeirat2016}.\\

\textbf{Acknowledgment} The author thanks Dr. Banu  and Sri Acharya Devanatha Swamiji
for the invitation to the Global Conference On Cosmology And Frontiers In Applied Astro Science held in February 7, 2017
at the Ethiraj College for Women in  Chennai, India. I also thank  Johanna Stachel and Friedel Thielemann for the useful discussions on various aspects of neutron stars and quark physics.

\end{document}